\input{psfig.sty}
\newcommand\cmmii {\rm \, cm^{-2}}
\newcommand\cmmiii {\rm \, cm^{-3}}

\newcommand\keV {\rm \, keV}

\newcommand\erg {\rm \, ergs}
\newcommand\ergsmi {\rm \, ergs \, s^{-1}}
\newcommand\ergcmmii {\rm \, ergs \, cm^{-2}}
\newcommand\ergcmmiismi {\rm \, ergs \, cm^{-2} \, s^{-1}}

\newcommand\s {\rm \, s}

\newcommand\G {\rm G}
\def\sn #1 #2 {#1 \times 10^{#2}}

\documentclass[12pt, preprint]{aastex}

\begin{document}

\title{An Extended Burst Tail from SGR 1900+14 with a Thermal X-ray Spectrum}

\author{
Geoffrey~T.~Lenters\altaffilmark{1},
Peter~M.~Woods\altaffilmark{3,4},
Johnathan~E.~Goupell\altaffilmark{2},
Chryssa~Kouveliotou\altaffilmark{3,4},
Ersin~{G\"o\u{g}\"u\c{s}}\altaffilmark{3,4},
Kevin~Hurley\altaffilmark{5}, 
Dmitry~Frederiks\altaffilmark{6},
Sergey~Golenetskii\altaffilmark{6}, and
Jean~Swank\altaffilmark{7}
}

\altaffiltext{1}{Calvin College, Department of Physics \& Astronomy, Grand
Rapids, MI  49546}
\altaffiltext{2}{Hope College, Department of Physics \& Engineering, Holland, MI
49422}
\altaffiltext{3}{Universities Space Research Association}
\altaffiltext{4}{NASA Marshall Space Flight Center, SD50, Huntsville, AL
35812}
\altaffiltext{5}{University of California, Berkeley, Space Sciences Laboratory,
Berkeley, CA 94720$-$7450}
\altaffiltext{6}{Ioffe Physical-Technical Institute, St.
Petersburg, 194021, Russia}
\altaffiltext{7}{NASA Goddard Space Flight Center, Greenbelt, MD 20771}

\begin{abstract}

The Soft Gamma Repeater, SGR 1900+14, entered a new phase of activity in April
2001 initiated by the intermediate flare recorded on April 18.  Ten days
following this flare, we discovered an abrupt increase in the source flux
between consecutive {\it Rossi X-Ray Timing Explorer} {\it RXTE} orbits.  This
X-ray flux excess decayed over the next several minutes and was subsequently
linked to a high fluence burst from SGR 1900+14 recorded by other spacecraft
({\it Ulysses} and {\it Wind}/Konus) while the SGR was Earth-occulted for {\it
RXTE}.  We present here spectral and temporal analysis of both the burst of 28
April and the long X-ray tail following it. We find strong evidence of an
exclusively thermal X-ray tail in this event and bring this evidence to bear on
other bursts and flares from SGR 1900+14 that have shown extended X-ray
excesses (e.g.\ 1998 August 29). We include in this comparison a discussion of
the physical origins of SGR bursts and extended X-ray tails.

\end{abstract}


\newpage

\section{Introduction}

Soft Gamma Repeaters (SGRs) are a small class of astrophysical objects
(currently four confirmed sources and one candidate source [Hurley 2000])
discovered by their emission of bright bursts of soft gamma-rays (with peak
luminosities ranging between $10^{38} - 10^{44} \ergsmi$). The SGR quiescent
counterparts have X-ray luminosities $\sim 10^{34} \ergsmi$ and their energy
spectra are generally well-fitted with a power-law model (index $\sim -2.2$)
attenuated by interstellar absorption; however, SGR~1900$+$14 has shown clear
evidence for an additional blackbody component with $kT \sim 0.5$ keV (Woods et
al.\ 1999a, 2001; Kouveliotou et al. 2001). Coherent pulsations have been
detected in the persistent emission of two sources (SGR~1806$-$20 at 7.5 s
[Kouveliotou et al.\ 1998] and SGR~1900$+$14 at 5.2 s [Hurley et al.\ 1999b])
with a spindown rate of $\sim$10$^{-10}$ s s$^{-1}$ (Kouveliotou et al.\ 1998 
and 1999). The Galactic SGRs (three) are located very close to
the plane of the Galaxy, indicating that SGRs belong to a young star
population. Furthermore, two of these SGRs may be associated with clusters of
very massive stars (Fuchs et al.\ 1999; Vrba et al.\ 2000; Eikenberry et al.\
2001), strengthening the claim that these are young objects.

The burst repetition timescale for these sources varies from seconds to
decades, while the bursts themselves are typically brief (durations $\sim 0.1
\s$ [e.g.\ {G\"o\u{g}\"u\c{s}} et al. 2001]). In contrast to these common brief
bursts, two SGRs have each emitted one large long duration burst. These large
bursts, or ``giant flares'' as they are sometimes called, are distinguished by 
their extreme energies ($\sim10^{44}$ ergs), their hard spectra at the onset, 
and their coherent pulsations during the decaying tail (lasting several
minutes), reflecting the spin of the underlying neutron star (Mazets et al.
1979; Hurley et al. 1999a; Feroci et al.\ 2001).

It is generally accepted today that SGRs are young, isolated, magnetized
neutron stars. Both magnetic and gravitational (i.e.\ accretion) energy have
been proposed as the source of the burst and persistent emission; however, the
accretion models have difficulty explaining several properties of the burst
emission (Thompson et al.\ 2000) as well as the lack of optical counterparts,
particularly in the case of SGR~0526$-$66 (Kaplan et al.\ 2001). This source is
located outside the Galactic plane and has a relatively small column density in
its direction, while the other three confirmed sources all lie near the
Galactic plane, where optical extinction is significant. A more convincing
model is that the SGRs are magnetars, i.e., highly magnetized neutron stars
(B$\sim 10^{14} - 10^{15} \G$) where magnetism is the largest source of free
energy in the system (Duncan \& Thompson 1992). In the magnetar model, the
strong magnetic field of the neutron star powers the persistent emission
through low level seismic activity and heating of the stellar interior
(Thompson \& Duncan 1996) as well as the burst emission through large-scale
crust fractures driven by an evolving magnetic field  (Thompson \& Duncan
1995). The super-Eddington burst fluxes are possible in the presence of such a
strong field because of the suppression of electron scattering cross sections
for some polarizations (Paczy\'nski 1992).

In this paper we focus on SGR~1900$+$14 which has been the most prolific SGR in
the last three years. SGR~1900$+$14 is the source of one of the giant flares
that was recorded on 1998 August 27 (Hurley et al.\ 1999a; Mazets et al.\
1999; Feroci et al.\ 1999, 2001).  Following the flare, the persistent X-ray
flux of SGR~1900$+$14 increased dramatically and decayed over the next $\sim$40
days as a power-law in time with an index $\sim-$0.7 (Woods et al.\ 2001). 
Nineteen days into the tail of this event, pointed X-ray observations revealed
that the enhancement in flux was contained within the non-thermal component of
the spectrum.  In addition to  this event, two other high-fluence events have
shown extended X-ray tails directly following the burst. These two events were
recorded on 1998 August 29 (Ibrahim et al.\ 2001) and more recently on 2001
April 18 (Guidorzi et al.\ 2001;  Kouveliotou et al.\ 2001; Feroci et al.\
2002; Woods et al.\ 2003).

Here we present results on the spectral evolution and temporal decay of the
X-ray tail of a burst recorded on 2001 April 28, ten days after the intense
April 18 event. Using data acquired with the {\it Rossi X-ray Timing Explorer}
(RXTE) Proportional Counter Array (PCA), we fit several spectral models to the
tail flux to constrain the radiative mechanism that governs its emission. We
further study in detail the evolution of the source pulse properties (pulse
shape and fraction) during this decay and compare the energetics of all four
events from SGR 1900$+$14 with extended tails. We discuss the implications of
these results on the current models for SGR emissions.

\section{2001 April 28 Burst}

The reactivation of SGR 1900$+$14 on 2001 April 18 triggered a series of
observations of the source with several satellites ({\it BeppoSAX}, {\it
RXTE}/PCA, {\it Chandra}). We observed the source with the {\it RXTE}/PCA
several times during the subsequent two weeks and discovered an intriguing
discontinuity in the source flux between consecutive {\it RXTE} orbits on April
28 (see Fig. 1). This trend was a strong indication of burst activity during
earth occultation of the source, since it was similar to the one observed
during the 1998 August 29 event (Ibrahim et al.\ 2001), in which the decaying
tail extended over multiple {\it RXTE} orbits. A search in {\it Ulysses} and
{\it Wind}/Konus data revealed an intense burst on 2001 April 28, peaking
approximately 121 seconds {\it before} our {\it RXTE} observations of the
source resumed at a higher flux level. The Fourier spectra of the {\it RXTE}
data provided an additional confirmation of the origin of the emission; the
spin period of SGR~$1900+14$ is prevalent during the tail orbit, while
undetectable in the previous (quiescent) orbit.

\subsection{Gamma-ray Observations}

On 2001 April 28, {\it Ulysses} and {\it Wind}/Konus observed a burst whose 
triangulated position (an annulus) contained SGR~1900$+$14. Its time history
was unusual  in that it had a duration of $\sim$2 s (Fig. 1), compared to the
more common, short duration SGR bursts. The {\it KONUS} energy spectrum was
well fitted by a power-law times an exponential function ($dN/dE \propto
E^{-0.5} {\exp}(-E/kT)$) with $kT=21.2$ keV. We estimated the ($25-100
\keV$) source fluence and peak flux (over 0.0625 s) to be $8.7 \times 10^{-6}$
ergs cm$^{-2}$ and $6.5 \times 10^{-5}$~ergs~cm$^{-2}$~s$^{-1}$, respectively.
The corresponding burst energy is $\sim \sn 2.0 41 $ ergs (assuming a source
distance for SGR 1900$+$14 of 14 kpc; Vrba et al. 2000).

\vspace{-0.4truein}
\begin{figure}[!htb]
\centerline{
\psfig{file=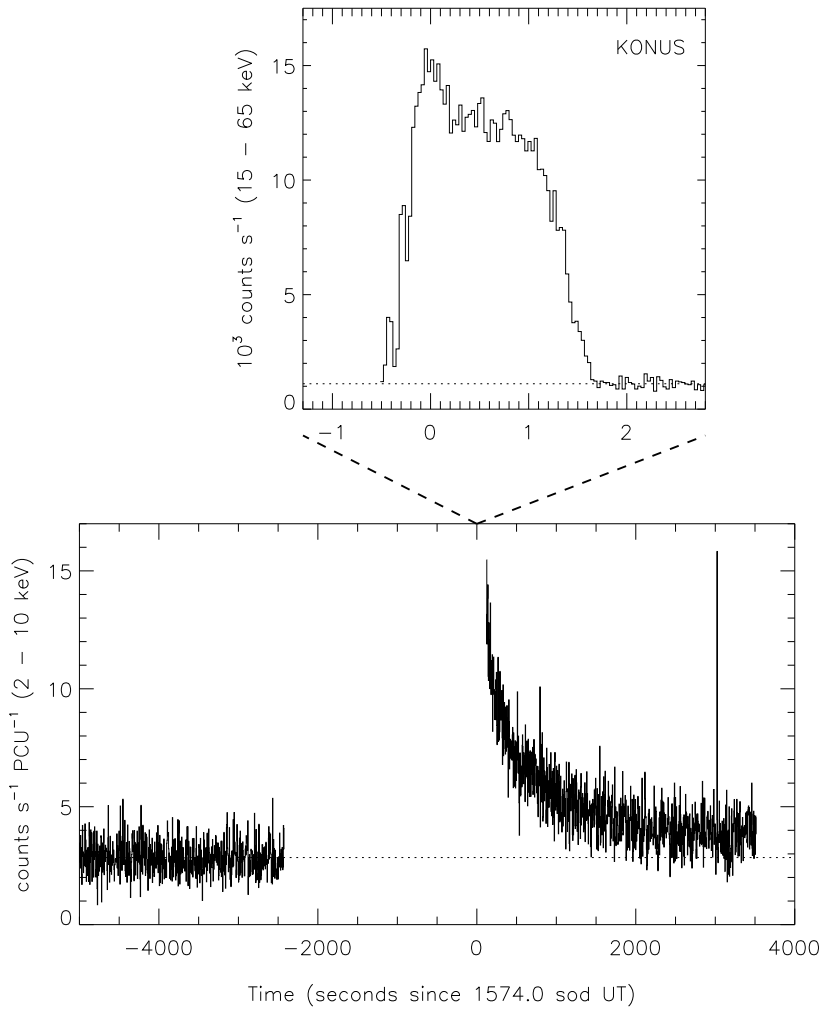,height=6.0in}}
\vspace{-0.2in}

\caption{{\it Top:} Konus 25-100 keV time history of the 2001 
April 28 burst from SGR 1900+14. The background level is indicated with a
dotted line. {\it Bottom:} {\it RXTE}/PCA 2-40 keV lightcurve before and
after the burst. The dotted line denotes the nominal background. }

\vspace{11pt}
\end{figure}

\subsection{X-ray Observations}

\subsection{Spectral Analysis}

We removed all bursts, the Galactic ridge component, quiescent SGR flux, and
other discrete X-ray sources for the spectral analysis of the X-ray tail
emission as follows. We first subtracted the instrumental and {\it RXTE}
orbital background from the source spectrum of the two orbits prior to the
burst using PCABACKEST (see  http://heassarc.gsfc.nasa.gov/
and links therein).  We then fitted the residual spectrum (2$-$30 keV), which is
dominated by the Galactic ridge emission (Valinia \& Marshall 1998), with a
two-component function (power-law plus Gaussian iron line). The width of the
iron line was fixed at 0.4 keV, while the other parameters were allowed to vary
in this fit. Using this model, we obtained a good fit to the data
($\chi^2_{\nu} =$ 1.17 for 52 degrees of freedom [dof]; see Table 1 for fit
parameters).  Please note that this spectral fit is {\it not} entirely the
quiescent spectrum of the SGR, but is rather the sum of the SGR quiescent
spectrum with the Galactic ridge and other discrete X-ray sources in the PCA
field-of-view.  This is a key point as we are interested in fitting the
spectrum of the SGR tail emission and {\it not} the sum of the tail plus
quiescent SGR flux.

\vspace{0.3truecm}
\begin{figure}[!htb]
\centerline{
\psfig{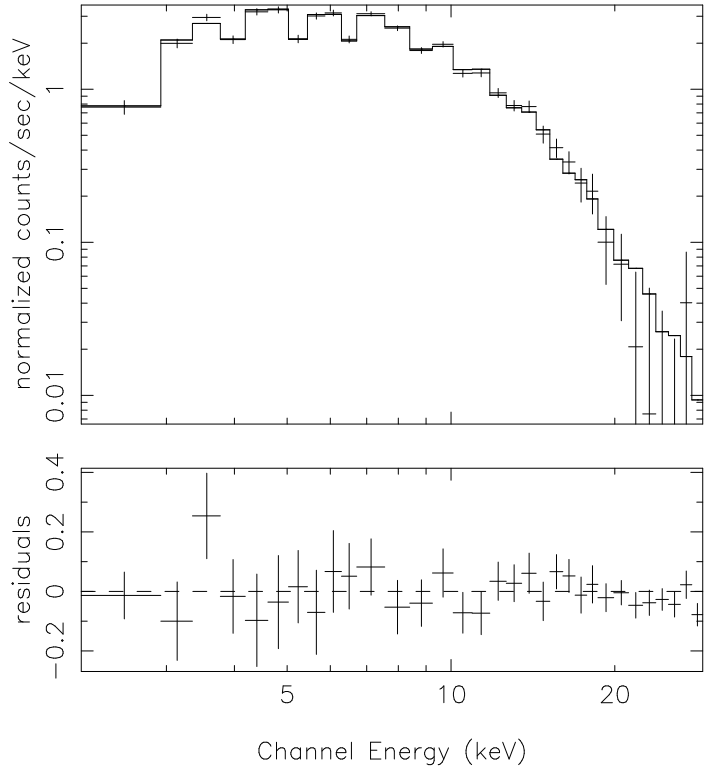}}

\caption{A 900 second exposure of the 28 April 2001 {\it RXTE}/PCA
spectrum with the attenuated (interstellar absorption) blackbody fit. }

\vspace{11pt}
\end{figure}

\begin{figure}[!htb]
\centerline{
\psfig{file=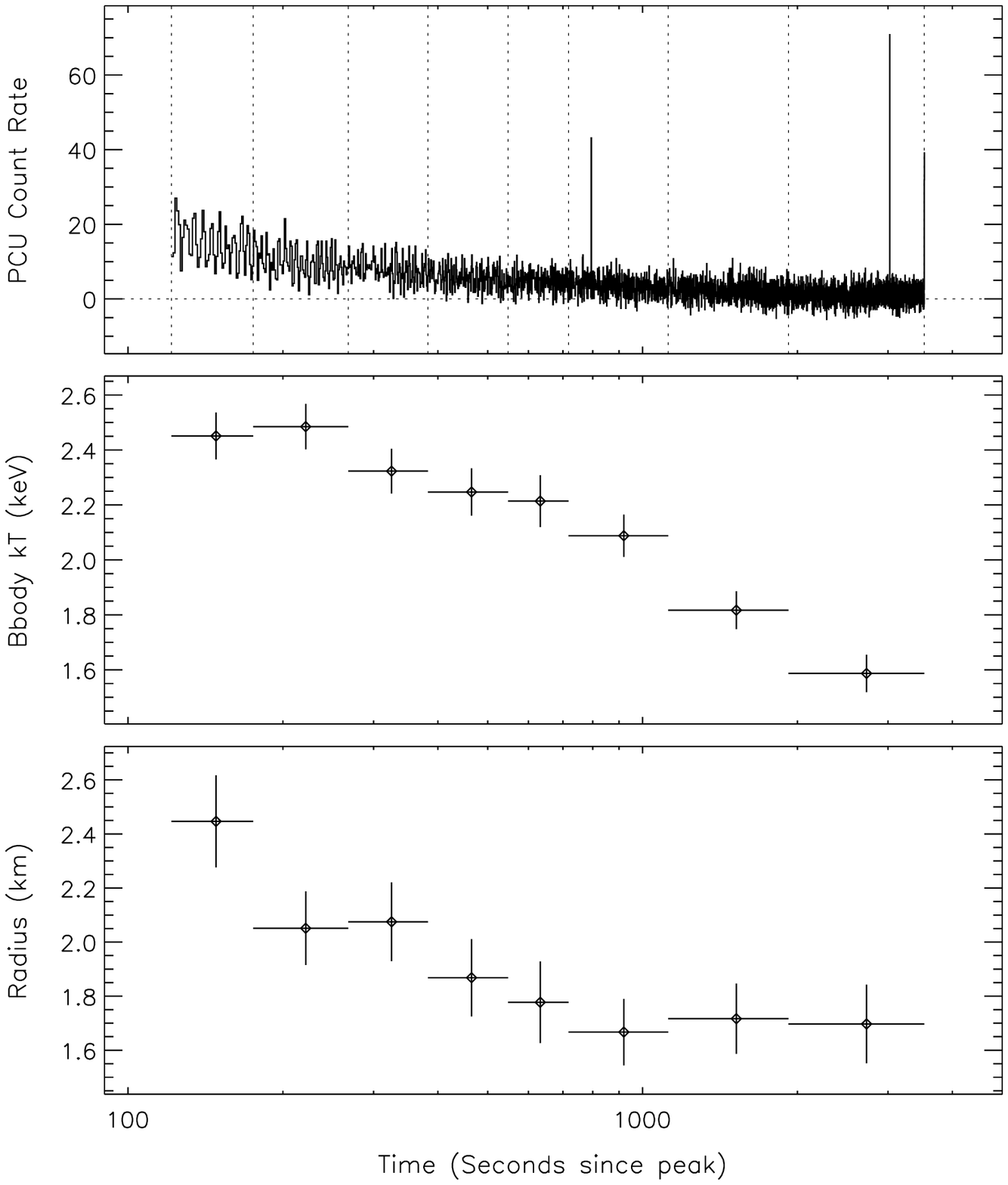,height=5.0in}}
\vspace{0.2in}

\caption{Parameters of the simple blackbody fit as a function of time for 
the 2001 April 28 decay. The top panel contains the lightcurve of the decay, 
the middle panel shows the blackbody temperature, and the bottom panel shows 
the radius of the emitting area. The vertical dotted lines in the top panel 
denote the time intervals over which spectra were fitted and the horizontal
dotted line is at zero counts s$^{-1}$. The radius was determined assuming a
distance  of 14 kpc to the source. }

\vspace{11pt}
\end{figure}

For the spectral analysis of the burst X-ray tail, we used the PCABACKEST
background together with the spectrum defined above as our ``total
background.''  More specifically, we defined a two part model, where one
component is the presumed tail spectrum and the second (fixed) part is the
remaining background as defined above (see Tables 1 and 2 for a summary of fit
parameters). We have fitted several spectral functions to the time-integrated
burst tail (900 s from the time the {\it RXTE} data start, see Fig. 2):
bremsstrahlung, power-law, and blackbody (each with interstellar absorption)
yielding $\chi^2$ values of 71.15, 119.7, and 21.24, respectively (29 dof for
each). Clearly the simple blackbody is the best fit to the data (0.85 chance
probability of measuring $\chi^2$ this large). We measure a temperature $kT =
2.35(5)$ keV, a radius $r = 1.74(9) $ km (assuming a 14 kpc distance) and an
effective hydrogen column density $N_H = 0.2(9) \times 10^{22} \cmmii$. We
attempted to fit two component models (blackbody plus power-law and
bremsstrahlung plus power-law) to the spectrum, but the bremsstrahlung plus
power-law failed to give a statistically acceptable fit to the data
($\chi^2_\nu = 2.7$) and the inclusion of the power-law component in the
blackbody plus power-law model was formally not required as evidenced by an
F-test (probability = 0.915). 

To study the spectral evolution of the tail, we divided the data into eight
intervals with nearly equal number of counts in each and fitted them with the
model determined above. We find that the column density ($N_H$) does not vary
significantly across the tail, while the temperature ($kT$) decreases
monotonically. Forcing the column density to be the same in each interval
results in an $N_H = 0.5(8) \times 10^{22} \cmmii$. The {\it RXTE}/PCA has low
sensitivity below 2 keV and therefore has difficulty measuring low column
densities. Hence, we fixed the column density at the value ($N_H = \sn 2.4 22
\cmmiii$) measured during {\it Chandra} observations (Kouveliotou et al. 2001)
that bracketed our PCA observations.  Both the temperature and normalization
(i.e.\ emitting area) decrease modestly through the tail (Fig. 3). 

Finally, we performed phase-resolved spectroscopy on the data during the tail. 
For the first $\sim$900 s of the tail (minus bursts), we accumulated spectra
for ten phase bins by folding the data on the spin ephermeris of the star
(Woods et al. 2003).  The background subtraction was handled in the same manner
as before. We performed a simultaneous fit to the ten phase binned spectra
using the blackbody model, again holding the $N_H$ fixed.  We find a marginally
significant (1.5 $\times 10^{-3}$ chance probability) variation in $kT$ versus
pulse phase (Fig. 4). 

\begin{figure}[!htb]
\centerline{
\psfig{file=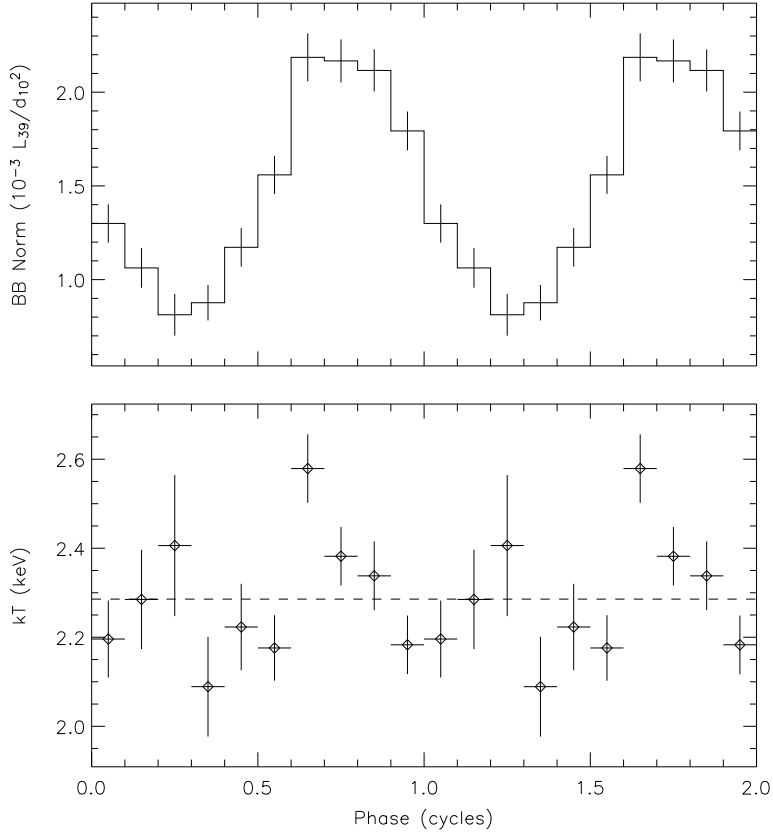,height=5.0in}}
\vspace{0.2in}

\caption{ {\it Top:} Blackbody normalization as a function of phase for
for the 2001 April 28 decay. {\it Bottom:} Blackbody temperature as a
function of phase for the 2001 April 28 decay. }

\vspace{11pt}
\end{figure}

\subsection{Temporal Analysis}

Using the blackbody spectral fit defined above, we measured the flux in four
energy bands (2$-$5 keV, 5$-$10 keV, 10$-$20 keV, and 2$-$20) to characterize
the flux decay as well as to determine the total fluence contained in the
extended tail emission. The errors in the flux are underestimated, since we
equate the relative error in the count rate to the relative error in the flux.
We fitted a power-law ($F \propto t^{-\alpha}$), a broken power-law ($F \propto
t^{-\alpha_1}$ for $t<t_b$ and $F \propto t^{-\alpha_2}$ for $t>t_b$), and a
power-law times exponential, [$F \propto t^{-\alpha} \exp(-{t/\tau})$]. In each
case $\alpha$ is the power-law decay index and $F$ is the flux. The fit
parameter $t_b$ in the broken power-law case is the time of the ``break'' in
the flux decay. We found that with the exception of the 2$-$5 keV energy band
(which was best fit by a single power-law) the decay was equally well fitted by
either power-law times exponential or broken power-law. We show the power-law
times exponential fit in Figure 5 and the power-law index and exponential decay
constant in Table 3 for each energy band. We find that the exponential decay
constant decreases with energy and the power-law index steepens with energy,
each indicative of spectral softening with time. Since the tail is still
detectable at the end of our observations, we quote a lower limit of
$\gtrsim$3.5 ksec for the tail duration. The power-law times exponential
function rapidly turns over after about 5 ksec and so integrating over longer
timescales does not contribute significantly to the fluence.  We thus integrate
the power-law times exponential function over 5 ksec and find the fluence to be
$\sn 3.0 -7 \ergcmmii$ in the 2$-$20 keV tail and $\sn 2.1 -7 \ergcmmii$ in the
2$-$10 keV tail.  The corresponding energies are $\sn 7.0 39 \erg$ and $\sn 4.9
39 \erg$ in the 2$-$20 keV and 2$-$10 keV energy band, respectively.

In addition to the temporal characteristics of the flux decay, we studied the
pulse properties of SGR~1900$+$14 during the X-ray tail.  The precise ephemeris
was determined elsewhere (Woods et al.\ 2003) from the complete {\it RXTE}/PCA
data taken during the 2001 April reactivation.  We folded seven time intervals
during the decaying burst tail as well as the two orbits prior to the burst for
comparison, all using 2$-$10 keV data.

\begin{figure}[!htb]
\centerline{
\psfig{file=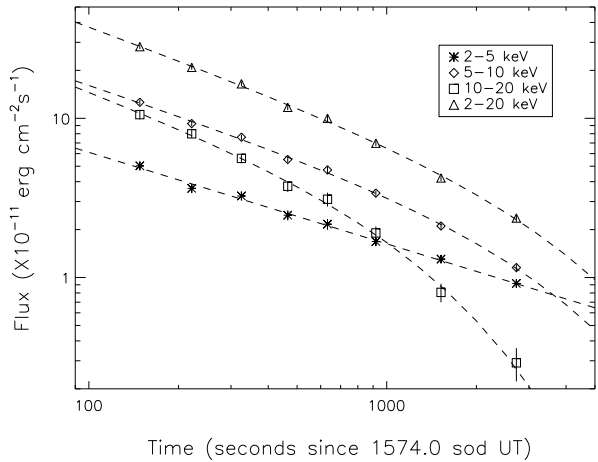,height=4.0in}}
\vspace{-0.7in}

\caption{Temporal decay of the flux from the 2001 April 28 burst from SGR
1900$+$14 in three energy bands. The dashed lines are the best-fit power-law
times exponential function to each band.}

\vspace{11pt}
\end{figure}

An accurate measure of the background level is required to measure the pulse
fraction of the SGR during this epoch.  Under normal circumstances, when the
SGR is in quiescence, we cannot use the PCA data to determine the absolute
level of the source flux because the contribution of the Galactic ridge is not
well known and thereby introduces a systematic error in the absolute background
determination that is comparable to the SGR count rate.  However, the PCA
cosmic background can be approximated during transient SGR flux enhancements
(e.g.\ following strong bursts) under certain assumptions: $i$) the PCA cosmic
background is constant over this brief time interval, $ii$) PCA data {\it
prior} to the flux increase are present, and $iii$) the pulse fraction of the
SGR directly before the burst is known.

\begin{figure}[!htb] 
\centerline{\psfig{file=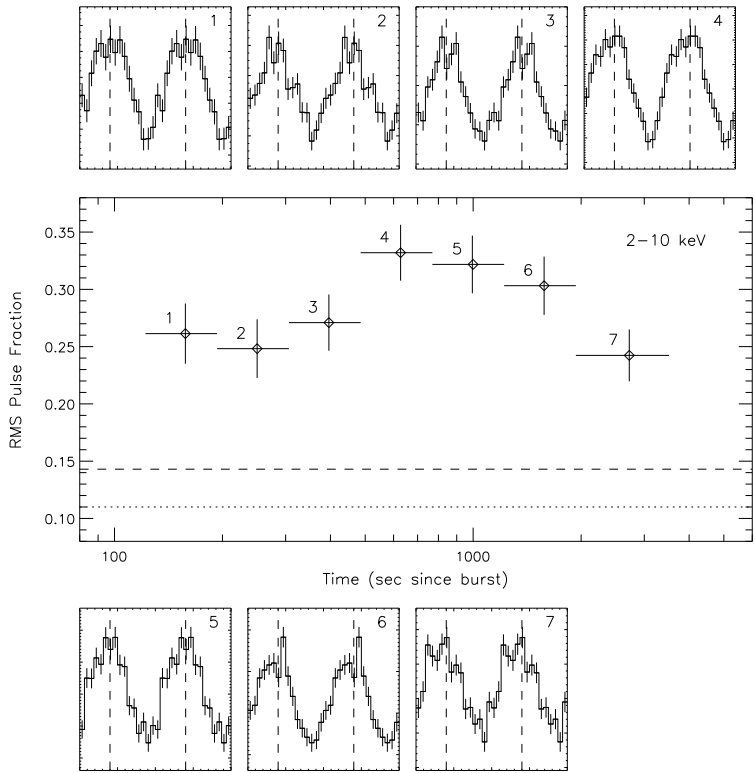,height=5.0in}}
\vspace{-0.5in}

\caption{Pulsed fraction  of SGR~1900$+$14 during the 2001 April 28 tail.
The horizontal dashed line denotes the interpolated pulse fraction prior to the
burst.  The dotted line marks the nominal quiescent pulse fraction.  The
vertical dashed lines in the pulse profiles are the phase of pulse maximum. The
pulse profiles shown cover 2 rotation cycles and have arbitrary flux units. 
The seven profiles correspond to the 7 numbered data points in the middle
panel.}

\vspace{11pt}
\end{figure}

{\it Chandra} and {\it BeppoSAX} observations indicated that the level of the
RMS pulse fraction (2$-$10 keV) increased directly after the April 18 flare to
$\sim$18\% and subsequently decayed to $\sim$14\% by May 1 (Kouveliotou et al.\
2001; Woods et al.\ 2003).  The pulse fraction is independent of energy in the
0.5$-$10 keV band during each of these epochs and the pulse shape does not
change significantly between epochs (Woods et al.\ 2003).  From these data, we
interpolated a pulse fraction of $\sim$14.3\% for the pre-April 28 burst PCA
data (i.e.\ the two orbits prior to the burst).  We used this pulse fraction to
determine the PCA cosmic background level during the pre-burst data on April 28
and assumed that this remains constant during the 0.2 day observation of the
tail. Using the constant cosmic background level we then measured the pulse
fraction during the tail of the burst. The RMS pulse fraction during the seven
intervals of the tail and the inferred level (dashed line) are shown in Figure
6.  We find that the pulse fraction increases significantly after the burst to
$\sim$33\%.  For completeness, we performed the same analysis assuming the
pre-burst pulse fraction was equal to the nominal quiescent value of $\sim$11\%
(Fig. 6, {\it dotted line}).  This systematically lowers the measured pulse
fractions by $\sim$1.5\% and thus does not change our results significantly.

We find no evidence for a systematic shift in phase of the pulse maximum or the
slope of the phases (i.e.\ pulse frequency) relative to the pre-defined
ephemeris, i.e., the pulse phase remained steady during the tail.  Figure 6
exhibits the pulse profiles for seven intervals used in this analysis; we find
that the shape is constant, with marginal changes (i.e., low-level phase
noise).

\begin{figure}[!htb]
\centerline{
\psfig{file=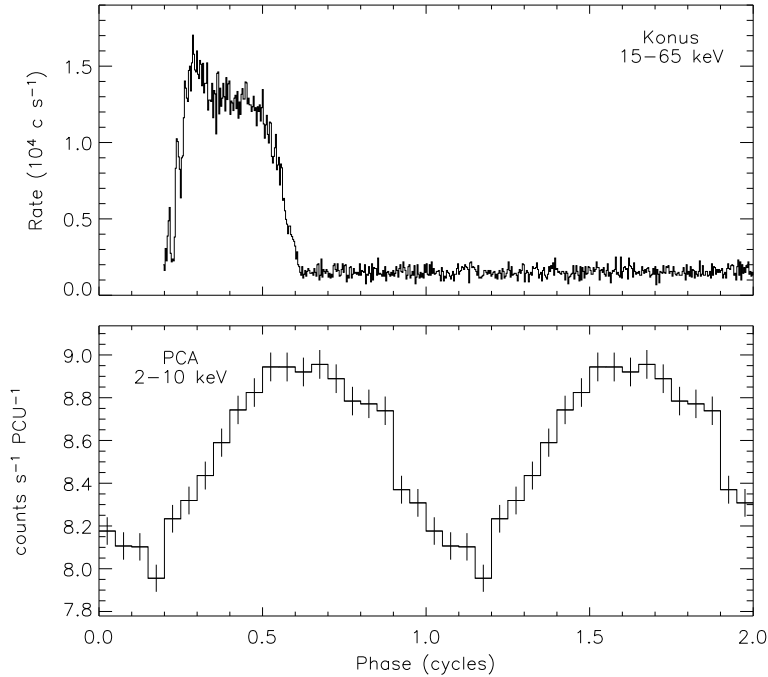,height=4.0in}}

\caption{Konus time history of the 2001 April 28 burst transformed to the
solar system barycenter.  The bottom panel shows the PCA pulse profile during
the time interval surrounding the burst.}

\vspace{11pt}
\end{figure}

Further, we investigated the phase occurrence of the burst itself.  We first
transformed the time history as observed with Konus to the solar system
barycenter (SSB) using mission specific tools.  This transformation was
confirmed by checking the alignment of an earlier burst observed from this
source on 1998 August 29 recorded with Konus, the Burst and Transient Source
Experiment (BATSE), and several other instruments (see \S3).  We have used the
BATSE time history for comparison because of its similar energy bandpass, good
time resolution and sensitivity.  The BATSE time history of the August 29 burst
was converted to the SSB using tools developed specifically for the BATSE
mission (Wilson et al.\ 1993).  A cross correlation between the BATSE and Konus
time histories shows that the relative transformation is accurate to within 14
msec for this event.  The absolute timing accuracy of Konus has been shown to
be better than 5 msec through extensive testing within the Interplanetary
Network.  We will adopt 14 msec as our error in temporal precision (or $\sn 2.7
-3 $ cycles in phase for SGR~1900$+$14) for the 2001 April 28 light curve.

The April 28 burst covers approximately 40\% of the SGR~1900$+$14 pulse cycle
(Fig. 7).  The burst begins near pulse minimum and terminates at pulse
maximum.  The combined timing analysis results (pulse fraction, pulse shape,
pulse phase, and burst phase) place significant constraints on the radiation
mechanism during the tail (see \S4).

\section{1998 August 29 Burst}

\begin{figure}[!htb]
\centerline{
\psfig{file=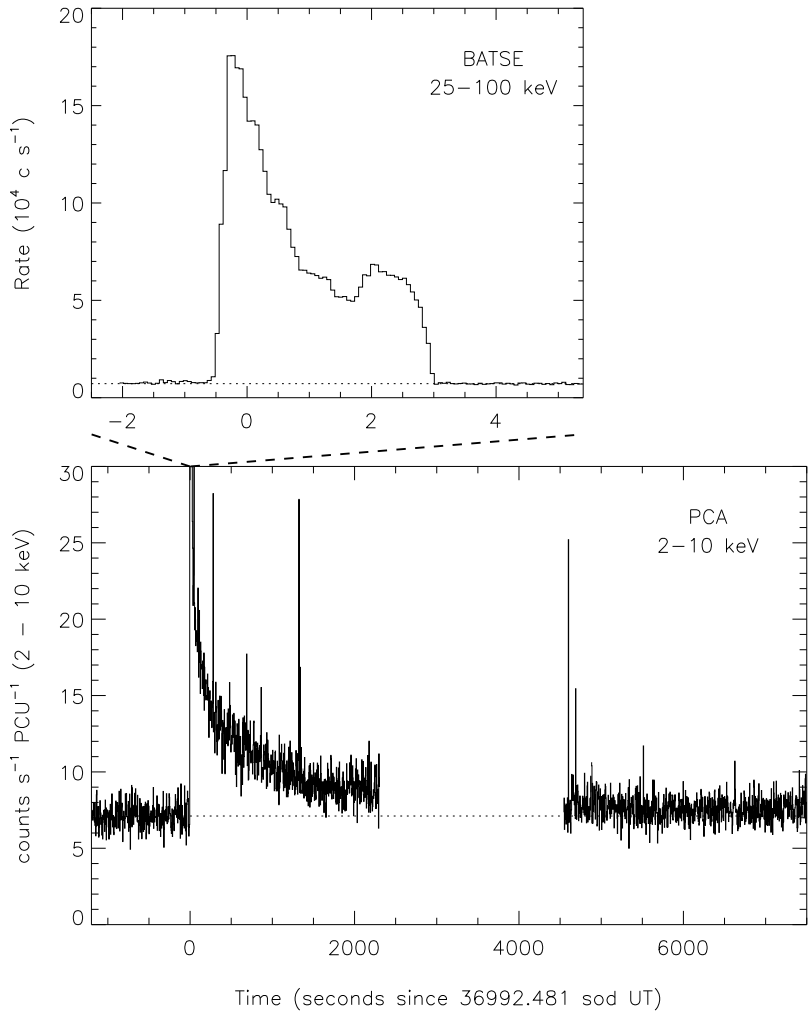,height=6.0in}}
\vspace{-0.25in}

\caption{{\it Top:} BATSE light curve of the 1998 August 29 burst
from SGR 1900$+$14. The background is denoted with the dotted line.  {\it
Bottom:} {\it RXTE}/PCA light curve before and after the burst. The
dotted line indicates the nominal background in both panels.}

\vspace{11pt}
\end{figure}

BATSE and {\it RXTE}/PCA observed  another relatively intense event from SGR
1900$+$14 on August 29, 1998 (Ibrahim et al. 2001). Similar to the 2001 April
28 event, it occurred in the days following a more energetic burst (on 27
August 1998) and exhibited a long duration X-ray tail.  Unlike the April event,
the August event had complete coverage with the {\it RXTE}/PCA (Fig. 8). To
compare the tail characteristics for both events, we analyze below the August
event using the same techniques that were implemented for the April event. We
have defined the `tail' emission onset as the time where the bulk of the burst
emission exhibits an abrupt decrease in flux ($\sim 3 \s$ after $t=0$ in Fig.
8). As discussed in Ibrahim et al. (2001), an abrupt change in the
bremsstrahlung temperature is seen at the same time.

\subsection{Spectral Analysis}

We first quantified the PCA cosmic background (including the Galactic ridge),
by fitting $\sim$1000 s of pre-burst data. As in the April 28 tail we find that
a power-law plus Gaussian iron line were sufficient to model the background
spectrum. The width of the iron line was again fixed at 0.4 keV, while the rest of
the parameters were allowed to vary in the fit (see Table 1). This model was
removed from the data (i.e.\ added as a fixed model) for all subsequent
spectral fitting of the August 29 X-ray tail.

For the August 29 event, we fitted bremsstrahlung, blackbody, power-law, and
blackbody plus power-law models, all having photoelectric absorption, to the
entire first orbit of the tail spectrum.  Consistent with the results of
Ibrahim et al.\ (2001), only the bremsstrahlung and blackbody plus power-law
models provided statistically acceptable fits to the data. In light of the
April 28 tail results we adopt the blackbody plus power-law model for this
analysis.  We measure a blackbody temperature $kT = 2.2(2)$ keV, a radius $r =
1.2(4) $ km (assuming a 14 kpc distance), a power-law photon index $\alpha =
2.1(2)$, and an effective hydrogen column density $N_H = 10(2) \times 10^{22}
\cmmii$. 

Following the method of analysis in the April event, we divided the data into
eleven segments with nearly equal counts, starting at the beginning of the
tail. Since the August event had complete coverage by the PCA, we were able to
look at four additional intervals earlier in the tail. We simultaneously fitted
all eleven intervals to a blackbody plus power-law fit and found the column
density consistent with being constant. We also find that the photon power-law
index is constant throughout the tail and thus we force the column density and
power-law index to be the same in each interval ($N_H = 10(2) \times 10^{22}
\cmmii$ and $\alpha = 2.3(3)$, respectively). The blackbody temperature shows a
stronger decay than the April 28 event; however, the emitting area shows first
a marginal decrease then a slight rise to a larger final area than beginning
area (see Fig. 9).  We note that the blackbody radius of the August 29 burst
tail is mildly correlated with power-law spectral parameters in the fit, unlike
the April 28 X-ray tail where no power-law component was detected.  Therefore,
the modest evolution we observe in the radius may be an artifact of this
cross-correlation and these small changes could possibly be attributed to
unmeasureable changes in the photon index.  However, the observed changes in
the blackbody temperature are too large to be accounted for from this effect. 
All of our spectral results on the August 29 burst tail are consistent with
those found by Ibrahim et al.\ (2001).

\begin{figure}[!htb]
\centerline{
\psfig{file=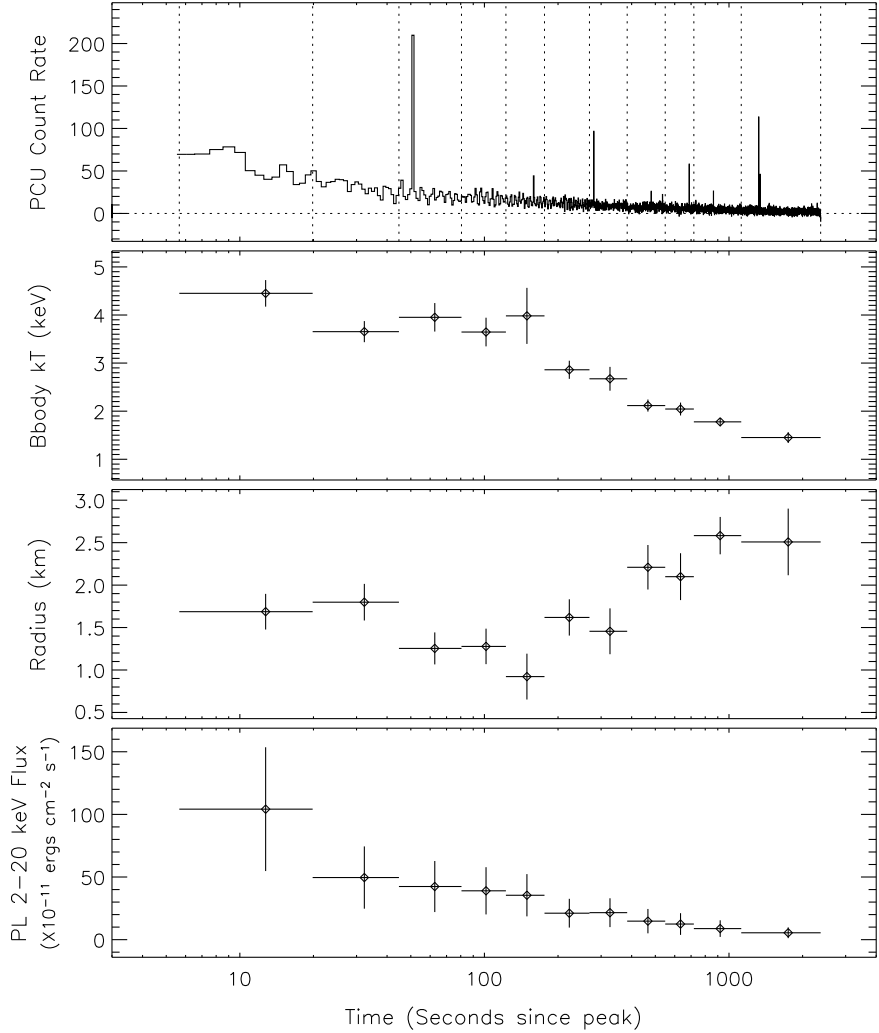,height=5.0in}}
\vspace{1.5cm}

\caption{Parameters of the blackbody plus power-law fit as a function of time
for the 1998 August 29 decay. The top panel contains the lightcurve of the
decay, the second panel shows the blackbody temperature, the third  panel shows
the radius of the emitting area, and the bottom panel shows the flux of the
power-law component. The vertical dotted lines in the top panel denote the time
intervals over which spectra were fitted and the horizontal dotted line is at
zero counts s$^{-1}$. The radius was determined assuming a distance of 14 kpc
to the source. }

\vspace{11pt}
\end{figure}

\subsection{Temporal Analysis}

\begin{figure}[!htb]
\centerline{
\psfig{file=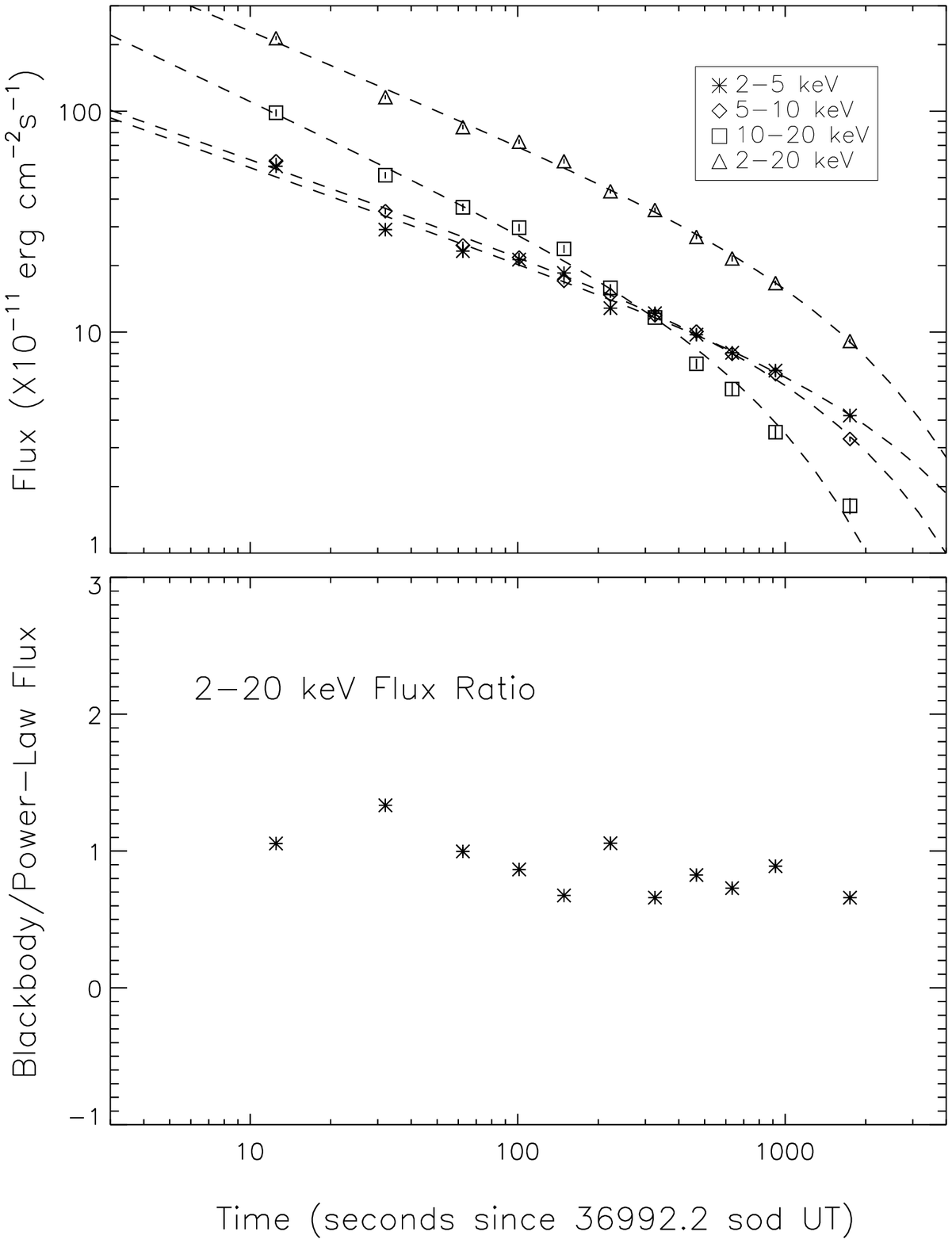,height=5.0in}}

\caption{Top panel shows temporal decay of the flux from the 1998 August 29
burst from SGR 1900$+$14 in three energy bands. The dashed lines indicate the
best-fit power-law times exponential function in each energy band. The bottom
panel shows the ratio of the power-law flux in the 2$-$10 keV energy band to
bolometric blackbody flux.}

\vspace{11pt}
\end{figure}

As in the April event we used the best-fit spectrum to determine the flux in
each time interval for four energy bands (2$-$5 keV, 5$-$10 keV, 10$-$20 keV,
and 2$-$20). Again, the errors in the flux are underestimated, since we equate
the relative error in the count rate to the relative error in the flux. We find
that for all energy bands the decay is best fit equally well by the broken
power-law and power-law times exponential. We show in Figure 10 the power-law
times exponential fit for each energy band with fit parameters given in Table
3. We also show in Figure 10, the ratio of the power-law flux in the 2$-$10 keV
energy band to bolometric blackbody flux. The ratio shows that at early times
the blackbody component dominates, while at later times the ratio increases to
around 1 consistent with the results of Marsden and White (2001). Similar to
the April 28 event the exponential decay constant decreases with energy and the
power-law index steepens with energy, indicating spectral softening with time. 

In the {\it RXTE} orbit directly following the August burst, the SGR count rate
(2$-$10 keV) remained significantly higher ($>5\sigma$) than the measured
pre-burst background.  During the subsequent orbit, the count rate was
consistent with the pre-burst background, thus we estimate the duration of this
tail to be $\sim$8 ksec, or the end of the last orbit in which the tail was
detected. Integrating the fit over 8 ksec yields a fluence and energy in the
2$-$20 keV tail of $\sn 6.4 -7 \ergcmmii$ and $\sn 1.4 40 \erg$, respectively,
and in the 2$-$10 keV tail $\sn 4.8 -7 \ergcmmii$ and $\sn 1.1 40 \erg$,
respectively. 

We then measured the pulse fraction for the August 29 event using the same
technique as before. Unfortunately, we do not have pulse fraction measurements
closely bracketing this event; the closest reliable RMS pulse fraction was
measured with BeppoSAX at 12\% on September 15-16 1998 (2$-$10 keV). Using this
value as the ``background'', we find for eight intervals during the tail that
the pulse fraction increased to a maximum of $\sim$20\% at $\sim 200 \s$ after
the burst peak and returned to the pre-burst value after several thousand
seconds (Fig. 11).  If we vary the assumed pre-burst pulse fraction to 8\%
and 16\%, the resulting maximum pulse fracion values measured during the tail
change to $\sim$17\% and $\sim$22\%, respectively.  Independent of the assumed
value of the pre-burst pulse fraction the increase of the pulse fraction from
pre-burst to post-burst remains highly significant.

Pulse shape changes are more prevalent during the tail of this event. We note
that the centroid of the primary peak is shifted in phase and the width of the
pulse broadens relative to the pre-burst pulse profile during the first
$\sim$30 s. After the first thirty seconds, the phase and shape of the profile
returns to the pre-burst profile as the pulse fraction rises.  Thereafter, the
phase of the average profile remains steady, although some low-level noise in
the shape of the profile persists.

\begin{figure}[!htb]
\centerline{
\psfig{file=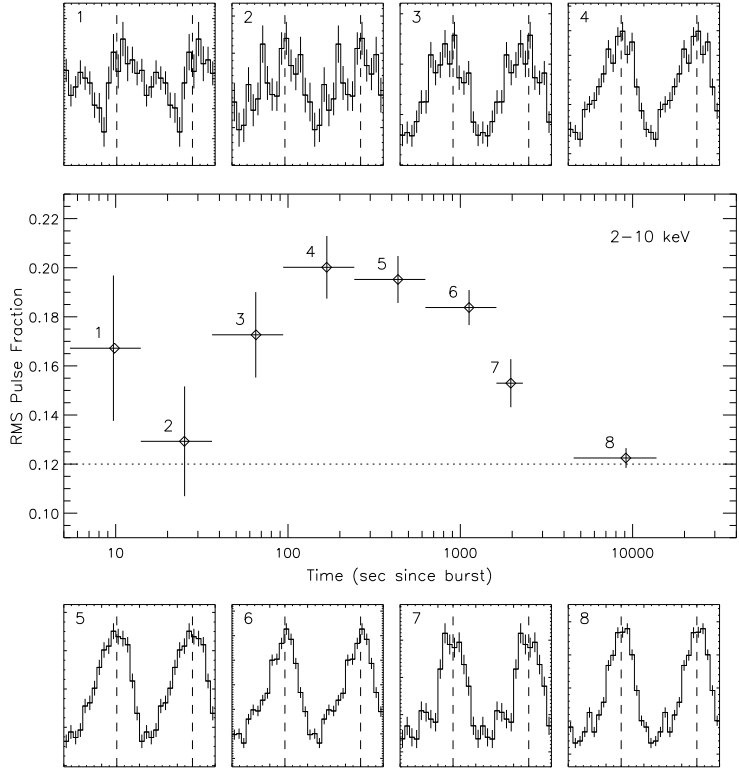,height=5.0in}}
\vspace{-0.55in}

\caption{Pulsed fraction in the 1998 August 29 decay. The horizontal
dotted  line indicates the pulse fraction prior to the burst. The vertical
dashed lines in the pulse profiles are the phase of pulse maximum.  The eight
profiles correspond to the 8 numbered data points in the middle panel.}

\vspace{11pt}
\end{figure}

We next investigated the phase occurrence of the August 29 burst.  The time
history of the August 29 burst as observed with BATSE was transformed to the
SSB and converted to pulse phase for SGR~1900$+$14 according to the spin
ephemeris found from previous work (Woods et al.\ 1999b).  The BATSE light curve
and PCA (2$-$10 keV) folded pulse profile from August 29 (minus bursts) are
shown in Figure 12.  The burst peak at gamma-ray energies lags slightly behind
the centroid of the X-ray pulse maximum ($\sim$0.1 cycle), although the onset
of the primary burst emission coincides with pulse maximum to within the
errors.

\begin{figure}[!htb]
\centerline{
\psfig{file=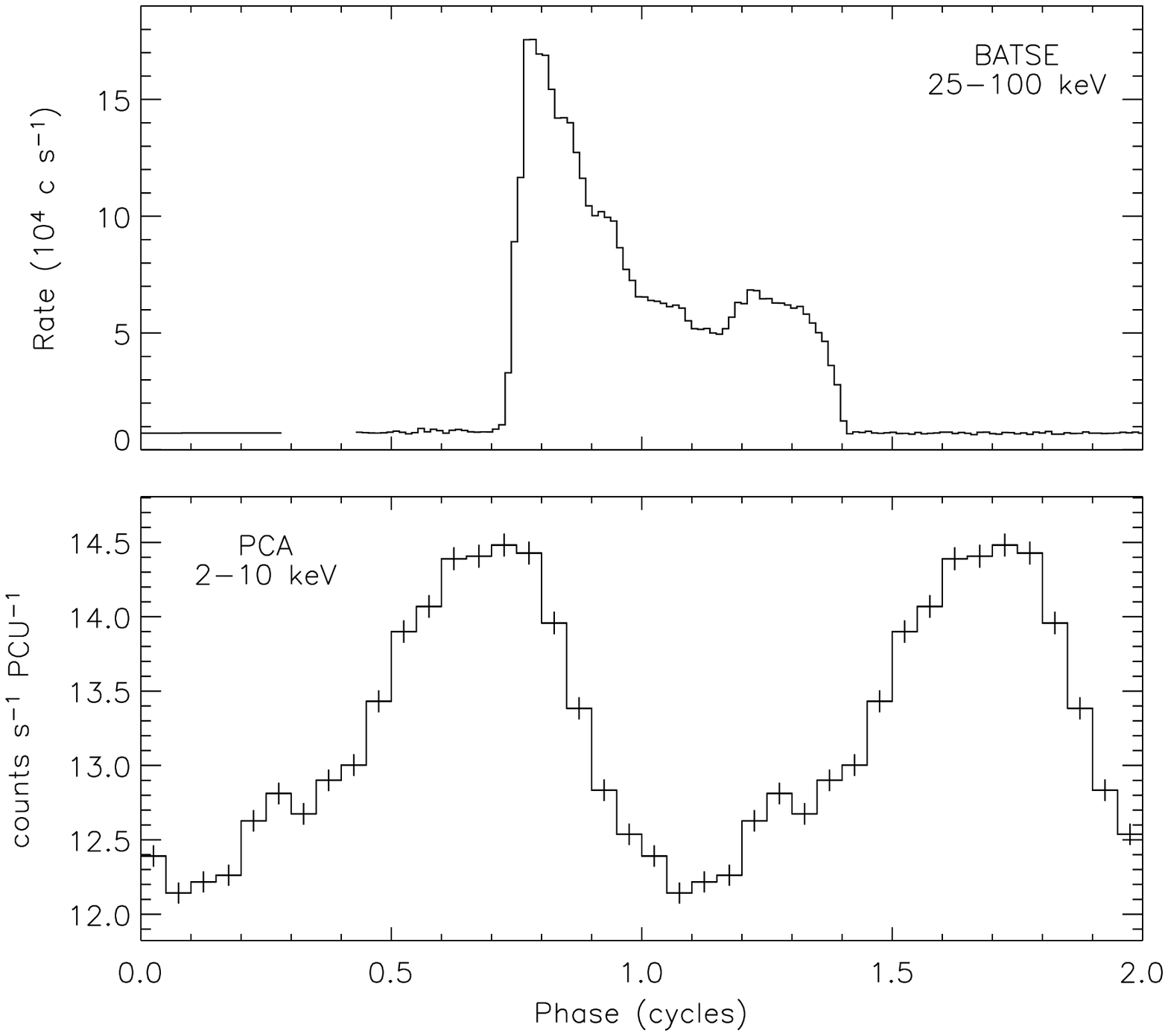,height=4.0in}}

\caption{BATSE time history of the 1998 August 29 burst transformed to the
solar system barycenter.  The bottom panel shows the pulse profile during the
time interval surrounding the burst.}

\vspace{11pt}
\end{figure}

\section{Burst Comparison}

The 1998 August 29 and 2001 April 28 bursts occur under very similar
conditions and originate at the same source, SGR 1900$+$14. They both occur
within days of much larger bursts or flares (1998 August 27 and 2001 April 
18). We therefore searched for similarities in temporal, spectral, and pulse 
properties between these two events. We have shown that the X-ray tails in 
both cases demonstrate spectral softening as the flux decays. The RMS pulse
fraction in both events was shown to rise by about a factor $\sim$2$-$3 during
the tails of each event while the phase of the pulsations remained more or less
steady.  Finally, the bursts themselves cover separate regions of pulse phase. 
The start times of these two bursts in gamma rays are $\sim$180 degrees out
of phase from one another. 

Since the August 29 event showed clear evidence of a power-law component, we
investigated whether the absence of the power-law component in the April 28
burst tail was due to lack of sensitivity or a significantly reduced relative
contribution from a hypothetical power-law component. If the April 28 burst
tail contained the same relative contribution of power-law to total flux as the
August 29 burst, then the expected power-law flux in the time integrated 2$-$20
keV April 28 tail would be $\sn 1.5 -10 \ergcmmiismi$. We proceeded to fit the
time integrated April 28 burst tail to a blackbody plus power-law model. All
parameters were allowed to vary, except the power-law photon index, which was
fixed at the best fit value for the August 29 burst tail ($\alpha = 2.1$).
Fixing the photon index was required to obtain convergence in the fit. As
expected, the power-law flux was consistent with zero having an upper limit of
$\sn 8.0 -11 \ergcmmiismi$ (90 \% confidence), a factor of 2 lower than the
expected flux. We conclude that the April 28 burst tail is exclusively thermal
and intrinsically different than the August 29 burst tail. 

We have estimated the energies in the tails of both the 2001 April 28 burst and
1998 August 29 burst in the previous sections. We find that the ratio of energy
output in the tail versus burst energy are very similar for these two bursts
(0.024 and 0.025, respectively). In Table 4 we include energy estimates of the
X-ray tails following the 2001 April 18 (Feroci et al.\ 2003) and the 1998
August 27 flares from SGR~1900$+$14 as well and find similar tail to burst
energy output ratios.  For the April 18 flare, we have assumed the power-law
fit to the X-ray tail plus persistent emission when calculating the tail
energy.  Due to the poor temporal coverage for this event and the presence of a
bump in the tail decay, the tail energy of this event is poorly determined
(Feroci et al.\ 2003).

For the August 27 flare, the X-ray tail remains above the nominal persistent
emission level for $\sim$40 days (Woods et al.\ 2001).  Unfortunately, the
coverage of this tail is not as complete as it is for the August 29 and April
28 bursts.  Therefore, in order to estimate the tail energy from the limited
information available, we must assume the 2$-$10 keV spectrum and pulse
fraction remain constant throughout the decay of the tail.  Each of these
assumptions is consistent with the published work on this X-ray tail.  The 2-10
keV pulse fraction measured at 19 days into this tail was found to be
consistent with the quiescent pulse fraction of $\sim$11\% (Woods et al.
2001).  Furthermore, the 2$-$5/5$-$10 keV pulsed amplitude softness ratio of
SGR~1900$+$14 (a coarse measurement of the spectrum within 2$-$10 keV) during
this tail was found to show only modest ($\sim$25\%) variations
({G\"o\u{g}\"u\c{s}} et al.\ 2002).  To estimate the X-ray energy in the tail
of the August 27 flare, we integrated a power-law fit to the pulsed flux
measurements in the 2$-$10 keV band (Woods et al.\ 2001) between the end of the
burst as observed in $\gamma$-rays (Feroci et al.\ 2001) and 40 days following
the flare minus the persistent emission flux over the same time interval.  We
measure a tail fluence of $\sn 1.5 -4 \ergcmmii$ or an isotropic energy of $\sn
3.5 42 d^{2}_{14 {\rm kpc}}$ ergs.

\begin{figure}[!htb]
\centerline{
\psfig{file=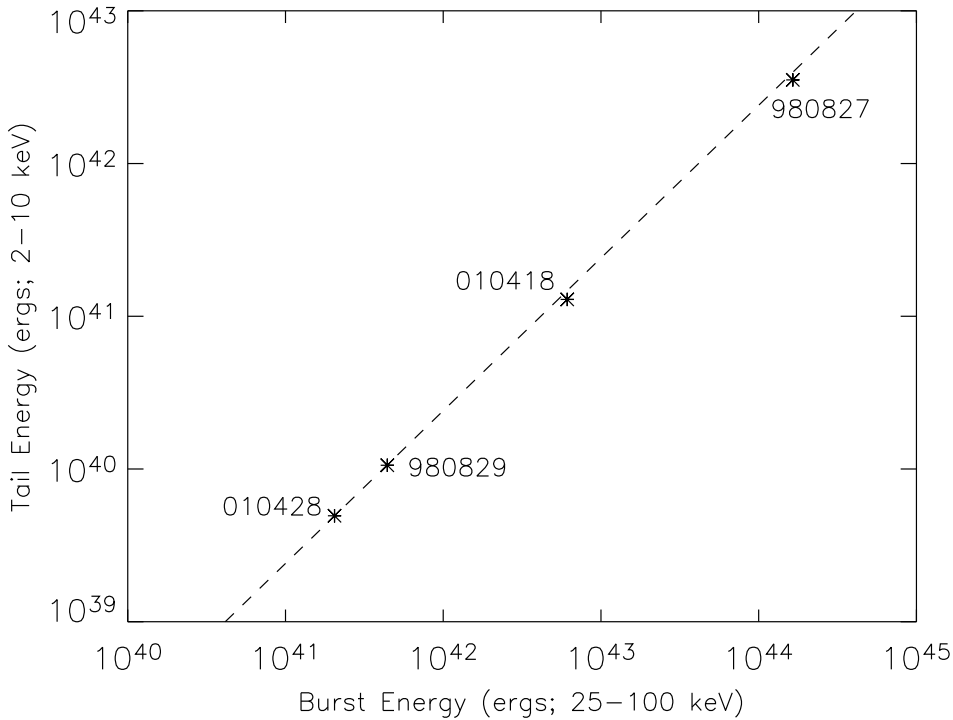,height=4.0in}}
\vspace{-0.in}

\caption{Tail energy versus burst energy output of four separate
bursts from SGR~1900$+$14.  An assumed distance of 14 kpc is used.}

\vspace{11pt}
\end{figure}

The burst and tail energies of the four tails are plotted in Figure 13.  The
data agree remarkably well with a constant tail to burst energy fraction (i.e.\
consistent with the dotted line which has a fixed slope of unity).  The average
ratio of tail to burst energy for these four bursts from SGR~1900$+$14 is
$\sim$2\%.  Note that this fraction reflects the ratio over finite, yet
consistent energy bands.

\section{Discussion}

We have detected, for the first time, a cooling, exclusively thermal X-ray tail
following a burst from SGR~1900$+$14 recorded on 2001 April 28.  During the
first $\sim$1 ks of the X-ray tail, we measure a 35\% decrease in the blackbody
temperature and a 30\% reduction in the emitting area. At all times, the
inferred emitting area encompasses a small fraction of the neutron star surface
assuming a 10 km radius neutron star.  We have confirmed earlier work (Ibrahim
et al.\ 2001) that showed a strong cooling blackbody present in the August 29
burst tail.  These observations of enhanced thermal emission during burst tails
provides strong evidence for burst induced surface heating in SGR~1900$+$14.

The pulse properties of these tails further constrain the location of the
heating on the stellar surface following the bursts. In quiescence,
SGR~1900$+$14 has a pulse fraction of 11~\%. In order to increase the pulse
fraction without changing the shape or phase of the pulse profile, one requires
that the additional emission that increases the pulse fraction come from the
same region of the stellar surface that is responsible for the peak of the
quiescent pulsations.  For the sake of argument, we will assume this region to
be the polar cap.  For example, suppose a region of the stellar surface were
heated and its location was at the same magnetic longitude, but  different
latitude than the polar cap.  For small offsets in latitude angle, this
scenario could account for a similar pulse profile, constancy of the pulse
phase from pre-burst to post-burst and, depending upon the observer's geometry,
an increase in pulse fraction.  For large offsets, all three observables will
be altered in a manner opposite to what is observed.  Alternatively, one could
have a small hotspot emerge at a different longitude, but the same latitude.
For small offsets in longitude angle this is compatible with observations. 
However, large offsets in longitude will present a phase lag (or lead) in the
pulse maximum, a change in pulse profile and a decrease in pulse fraction. 
Considering these two scenarios, we conclude that the observations restrict the
region of enhanced emission to be located within an ellipse surrounding the
polar cap.  Note that there are no stringent constraints on the altitude of the
excess emission, only the angular position relative to the magnetic axis.
Quantifying the restrictions in magnetic longitude and latitude is difficult as
it will depend upon the observer's geometry, gravitational light bending, beam
geometry, and so on.

We have shown that the August 29 and April 28 events began approximately 180
degrees apart in phase.  Earlier work by Palmer (2000) showed that the
occurrence in phase of bursts from SGR~1900$+$14 during late August and early
September 1998 were random.  This was true for all bursts as well as subsets of
bursts at different energies.  The random occurrence of bursts in pulse phase
indicates that the bursts themselves are not localized near the region
contributing to the pulsed emission; however, the ensuing flux enhancements
(well-correlated with burst activity [Woods et al. \ 2001]) indicate localized
heating near the putative polar cap. 

With these observational constraints in mind, we now consider the magnetar
model. In this model, it is postulated that the build up of magnetic stresses
within the neutron star crust are sufficient to overcome the tensile strength
of the crust and crack it (Thompson \& Duncan 1995).  Fracturing the stellar
crust perturbs field lines and ultimately leads to a trapped pair-photon
fireball.  Thompson \& Duncan (1995) considered the likelihood of heating of
the neutron star surface by the resultant trapped fireball suspended above the
fracture site.  For shallow heating, they estimated the cooling time scale to
be equivalent to the burst duration.  In this time scale estimate, however,
they did not account for the possibility of vertical expansion against gravity
in the atmosphere which could extend passive surface cooling up to $\sim$10$^3$
sec (Thompson, Lyutikov \& Kulkarni 2002). Lyubarsky, Eichler, \& Thompson
(2002) also showed that longer time scale cooling can also be achieved through
deep heating of the crust that leads to a longer conduction time of the heat to
the surface and that the temporal evolution of cooling after deep crustal
heating could resemble a power-law in time with a decay index $\sim-$0.7.  This
picture fits nicely with the cooling thermal components observed during these
X-ray tails.  

One requirement of this model with regards to these data is that the burst
fracture site be located at or very near one of the magnetic poles.  The
proximity of the fracture site (i.e.\ burst) to the polar cap vicinity is
required to satisfy the constraints introduced by the pulsed flux
observations.  For this model to work, however, the bursts themselves must be
visible at all rotational phases (i.e.\ even when the polar cap is facing
opposite the observer).  This suggests that the light emitted from the burst
originate at sufficiently high altitudes or must be scattered as has
been suggested for the persistent emission from these sources (Thompson et al.\
2002). However, the trapped fireball for most bursts is held close to the
surface and it is questionable whether strong scattering can occur at much
higher luminosities than the persistent X-ray luminosity and for higher photon
energies, particularly without substantial spectral variations in the bursts
with pulse phase.

An alternative to explain these observations would be that the bursts have a
magnetospheric origin (Lyutikov 2002).  Here, we consider a concentric sphere
surrounding the neutron star at a large radius ($R > 10 R_*$) where magnetic
reconnection is occurring and giving rise to SGR bursts.  The high altitude
allows bursts to be observed for nearly all rotational phases.  Moreover, the
field lines permeating this surface are preferentially concentrated at the
magnetic poles.  Hence, particles accelerated along these field lines will
impact the stellar surface at the polar cap regions.  This scenario we have
outlined is qualitatively compatible with the observations, although, further
quantitative modeling is needed with regards to surface heating/cooling by
ablation, burst timescales, burst spectra, etc. before it can be considered
further.

Finally, the observation that the ratio of tail to burst energy is consistent
with being constant between the four events from SGR~1900$+$14 is an intriguing
one.  However, we would like to stress that this result is based upon a small
number of events and some using predominantly inferred fluxes from the measured
pulsed intensity, therefore, we regard this currently as a tentative result. 
In addition to potentially confirming this result, the detection of several
more SGR burst tails will provide useful diagnostics for the burst mechanism as
shown from this work.  Assuming the observed conversion of burst to tail energy
extends to lower energy burst events, X-ray tails can be observed from the more
plentiful low-energy SGR bursts (down to $\sim$$\sn 2 39 d^{2}_{14 {\rm kpc}}$
ergs) with sensitive, low background X-ray telescopes such as {\it Chandra} and
{\it XMM-Newton}.

\acknowledgments{\noindent {\it Acknowledgments} -- We thank Marco Feroci for
useful comments and suggestions regarding the temporal fits to the X-ray tails.
GL acknowledges support from NASA's Michigan Space Grant Consortium seed grant
program, Hope College, and an ADP grant (NAG 5-11608). PMW and CK would like to
acknowledge support from NASA through grants NAG 5-11608 and NAG 5-9350. JG is
grateful for the Sherman Fairchild Foundation support of Hope College's REU
program. KH is grateful for Ulysses and Ineterplanetary Network support under
JPL Contract 958056 and NASA Grant NAG 5-11451.}

\newpage

\begin{center}
\begin{deluxetable}{l|ccccc}
\scriptsize
\tablecaption{Persistent Emission Parameters. \label{tbl-1}}
\tablewidth{6.5in}

\tablehead{
& \colhead{N$_{\rm H}$ (10$^{22}$~cm$^{-2}$)}   &  
\colhead{$\alpha$\tablenotemark{a}}          &  
\colhead{Emission Line\tablenotemark{b}}     &  
\colhead{Line Width\tablenotemark{b}} &  \colhead{$\chi^2$/dof} 
}

\startdata

28 April 2001 &      1.7~$\pm$~0.9     &  1.95~$\pm$~0.08  &  
6.62~$\pm$~0.08  &  0.4~(frozen)  &  60.6/52 \\

29 August 1998 &     3.1~$\pm$~0.4     &  2.72~$\pm$~0.05  &
6.4~$\pm$~0.3    &  0.4~(frozen)   &  58.1/70 \\

\enddata

\tablenotetext{a}{Power law photon index}
\tablenotetext{b}{Gaussian line profile, modeling Galactic ridge}
\end{deluxetable}
\end{center}

\begin{center}
\begin{deluxetable}{cccccc}
\scriptsize
\tablecaption{Summary of spectral fits. \label{tbl-2}}
\tablewidth{6.5in}

\tablehead{
\colhead{Model}    &  
\colhead{N$_{\rm H}$ (10$^{22}$~cm$^{-2}$)}   &  \colhead{$kT$ (keV)}     &  
\colhead{R$_{\rm bb}$ (km)\tablenotemark{a}}  &  
\colhead{$\alpha$\tablenotemark{b}}          &
\colhead{$\chi^2$/dof} 
}

\startdata

\multicolumn{6}{l}{28 April 2001} \\ 
BB          &     0.2~$\pm$~0.9     &     2.35~$\pm$~0.05  
            &     1.74~$\pm$~0.09   &     \ldots
            &     21.2/29 \\

BREMSS      &     13~$\pm$~1        &     10.7~$\pm$~0.9
            &     \ldots            &     \ldots     
            &     71.1/29 \\

PL          &     17~$\pm$~1        &     \ldots
            &     \ldots            &     2.26~$\pm$~0.08
            &     119.7/29 \\



\hline\hline

\multicolumn{6}{l}{29 August 1998} \\ 

BB$+$PL  &     10~$\pm$~2           &     2.2~$\pm$~0.2
         &     1.2~$\pm$~0.4        &     2.1~$\pm$~0.2
         &     30.7/35 \\

BREMSS   &     10.2~$\pm$~0.6       &     12.9~$\pm$~0.7
         &     \ldots               &     \ldots
         &     30.1/37 \\

PL       &     14.0~$\pm$~0.8       &     \ldots
         &     \ldots               &     2.16~$\pm$~0.05
         &     55.5/37 \\

\enddata

\tablenotetext{a}{Blackbody radius without general relativistic correction
(assumes d = 14 kpc) } 
\tablenotetext{b}{Power law photon index}

\end{deluxetable}
\end{center}

\begin{center}
\begin{deluxetable}{cccc}
\scriptsize
\tablecaption{Parameters of Flux Decay. \label{tbl-3}}
\tablewidth{6.5in}

\tablehead{
\colhead{Event}    &  
\colhead{Energy Band (keV)}   &  \colhead{Power-law Index}     &  
\colhead{Decay Constant (s)}  
}

\startdata

28 April 2001  &  2-5   &  -0.58~$\pm$~0.07  &  >8 $\times 10^3$ \\
               &  5-10  &  -0.62~$\pm$~0.04  &  4~$\pm$~1 $\times 10^3$ \\
               &  10-20 &  -0.7~$\pm$~0.1    &  1.5~$\pm$~0.4 $\times 10^3$ \\
	       &  2-10  &  -0.62~$\pm$~0.04  &  8~$\pm$~2 $\times 10^3$ \\
               &  2-20  &  -0.68~$\pm$~0.04  &  5~$\pm$~1 $\times 10^3$ \\
               
29 August 1998 &  2-5   &  -0.43~$\pm$~0.02  &  5~$\pm$~1 $\times 10^3$ \\
               &  5-10  &  -0.43~$\pm$~0.01  &  2.6~$\pm$~0.2 $\times 10^3$ \\
               &  10-20 &  -0.57~$\pm$~0.02  &  1.21~$\pm$~0.09 $\times 10^3$ \\
	       &  2-10  &  -0.43~$\pm$~0.01  &  3.5~$\pm$~0.4 $\times 10^3$ \\
               &  2-20  &  -0.510~$\pm$~0.008  &  2.9~$\pm$~0.2 $\times 10^3$ \\

\hline\hline

\enddata

\end{deluxetable}
\end{center}

\begin{center} 
\begin{deluxetable}{lcc}
\scriptsize
\tablecaption{Burst Fluences (25$-$100 keV) and Tail Fluences (2$-$10 keV). \label{tbl-4}}
\tablewidth{6.5in}

\tablehead{ \colhead{Event}    & 
\colhead{Tail Fluence ($\ergcmmii$)}   &  \colhead{Burst Fluence ($\ergcmmii$)}  }

\startdata

27 August 1998 &  $\sn 1.5 -4 $   &  $\sn 7.0 -3 $   \\
29 August 1998 &  $\sn 4.8 -7 $   &  $\sn 1.9 -5 $   \\
18 April 2001  &  $\sn 5.5 -6 $   &  $\sn 2.6 -4 $   \\
28 April 2001  &  $\sn 2.1 -7 $   &  $\sn 8.7 -6 $   \\

\hline\hline

\enddata

\end{deluxetable}
\end{center}

\end{document}